\documentclass[a4paper,11pt]{article}
\usepackage{pos}
\usepackage{lineno}

\title{First measurements of in-jet fragmentation and correlations of charmed mesons and baryons in pp collisions with ALICE}

\author*[]{Antonio Palasciano}

\onbehalf{for the ALICE collaboration}

\affiliation[]{University and INFN, Bari\\}

\emailAdd{antonio.palasciano@ba.infn.it}

\abstract{
Fragmentation functions are one of the key components of the factorisation theorem used to calculate heavy-flavour hadron production cross sections. The non-perturbative nature of fragmentation functions necessitates that they are constrained through experimental measurements, commonly performed in the clean environments of $\mathrm{e}^{+}\mathrm{e}^{-}$ and ep collisions. However, recent measurements of charm-hadron transverse-momentum spectra and the ratios of charmed-hadron abundances in pp collisions have questioned the universality of fragmentation functions between leptonic and hadronic collision systems in the baryon sector.
In this contribution, we present measurements of differential observables of heavy-flavour hadrons that also consider the hadronic density surrounding the hadron. These measurements provide additional information to the previously reported  baryon-to-meson results and allow us closer access to the charm fragmentation functions. We report the fraction of longitudinal momentum carried by $\mathrm{D^{0}}$ and $\mathrm{D^{+}_{s}}$ mesons as well as $\Lambda^{+}_\mathrm{c}$ baryons. We also report the azimuthal correlation distributions between heavy-flavour decay electrons and charged particles in pp and p--Pb collisions, as well as between $\Lambda^{+}_\mathrm{c}$ baryons and charged particles in pp collisions, which provide quantitative access to the angular profile, $p_{\mathrm{T}}$ and multiplicity distributions of the jets produced by the heavy-quark fragmentation.
}

\FullConference{%
  $11^\mathrm{th}$ International Conference on Hard and Electromagnetic Probes of High-Energy Nuclear Collisions - HP2023,\\
  26-31 March, 2023\\
  Aschaffenburg, Germany
}

\begin{document}
\maketitle

\section{Introduction}
Heavy-flavour (HF) measurements in ultra-relativistic hadronic collisions play a crucial role in probing and testing our current understanding of the perturbative Quantum Chromodynamics (pQCD). In this theoretical framework, the factorization theorem, used to compute HF hadron production cross-section, foresees the separation of perturbative quantities, as the partonic cross-sections, from non-perturbative quantities as the parton distribution functions (PDFs) and the fragmentation functions (FF), which are evaluated by exploiting measurements.
The recently observed enhancement of baryon-to-meson ratios with respect to measurements in $\mathrm{e^{+}e^{-}}$, $\mathrm{e^{-}p}$ collisions~\cite{D0Lc, D0Lcmult} has questioned the fragmentation universality across colliding systems, exposing discrepancies between experimental measurements in hadronic collisions and models relying on FF constrained to $\mathrm{e^{+}e^{-}}$ and $\mathrm{e^{-}p}$ data.
Charm-tagged jets measurements and azimuthal correlations between charm-origin particles and other charged particles, differently from single-particle studies, directly access complementary aspects of the charm fragmentation, providing a tool for the description of the charm shower. 

\section{Heavy-flavour tagged jets}
The study of charm-baryon production in jets can provide additional insights into the charm hadronisation mechanisms in pp collisions. Charm-tagged jets were identified by the presence of prompt charmed hadrons, such as D mesons or charmed baryons, among their constituents. In events where at least one charm hadron candidate was reconstructed, an anti-$k_\mathrm{T}$ clustering algorithm as implemented in the FastJet~\cite{fastjet} package was applied to the reconstructed charged particles to define the jet produced by the charm showering process. Corrections accounting for the D-meson reconstruction efficiency, acceptance, and beauty-hadron feed-down contribution were applied and an unfolding procedure was performed to account for detector effects such as non-reconstructed particles and limited resolution on the jet momentum.

\begin{figure}[t]
	\centering
	{\includegraphics[width = 0.75\textwidth]{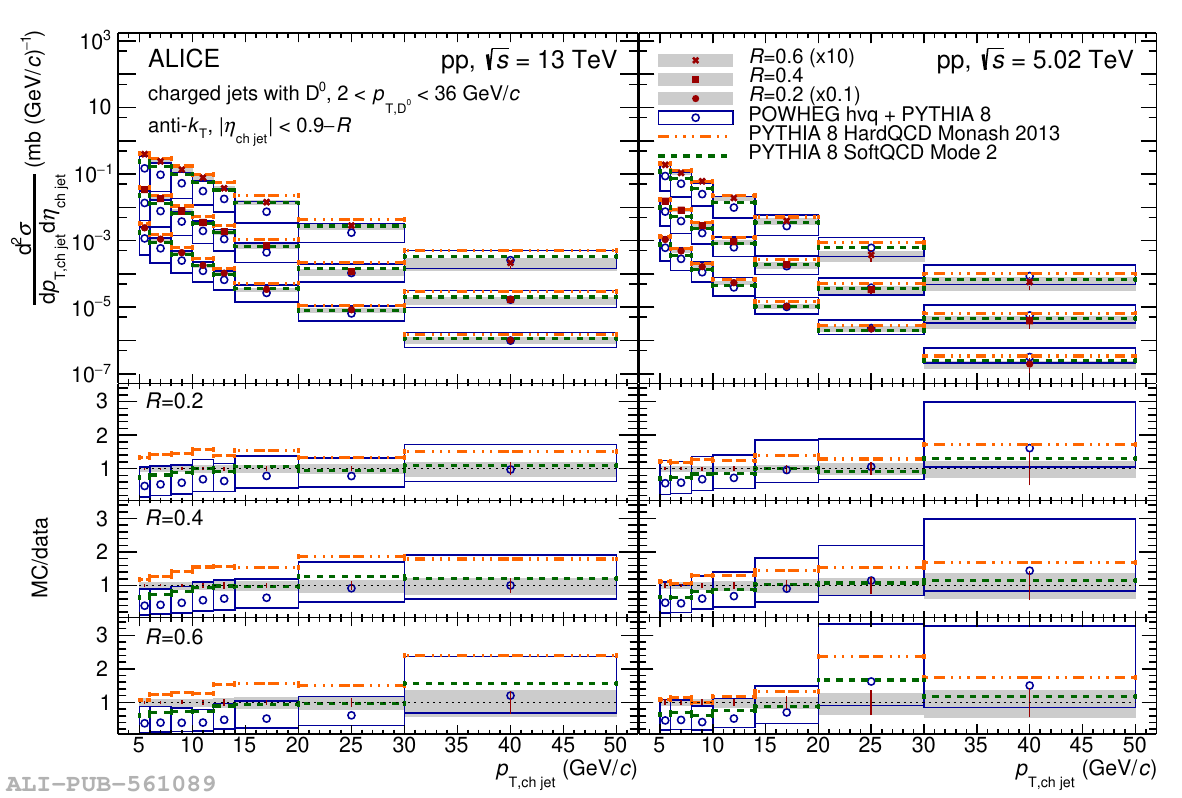}}
	\label{fig:JetProd}
    \caption{$\mathrm{D^0}$-jet production cross-section measurements in pp collisions at $\sqrt{s} =$ 5.02 and 13 TeV for three different resolution parameter $R$ = 0.2, 0.4, 0.6, compared to POWHEG + PYTHIA 8 and PYTHIA 8 Monte Carlo predictions.}
    \label{fig:XSectionD}
\end{figure}

The ALICE Collaboration has measured the $p_\mathrm{T}$-differential cross section of $\mathrm{D^0}$-tagged charm jets~\cite{d0jets}. In Fig.~\ref{fig:XSectionD}, the results for three different values of the jet resolution parameter $R$ are compared to PYTHIA 8 simulations with Monash-2013 tune and with color reconnection beyond-leading-colour (CR-BLC) Mode 2~\cite{pythia, colorReconnection} tune, as well as with POWHEG + PYTHIA 8~\cite{powheg} simulations, in pp collisions at $\sqrt{s}=$ 5.02 and 13 TeV. The two PYTHIA 8 predictions correctly reproduce the measured $p^\mathrm{ch~jet}_\mathrm{T}$ -differential cross sections for both collision energies.
A good agreement within the experimental and theoretical uncertainties is also observed in POWHEG + PYTHIA 8 predictions.

\begin{figure}[t]
	\centering
	{\includegraphics[width = 0.47\textwidth]{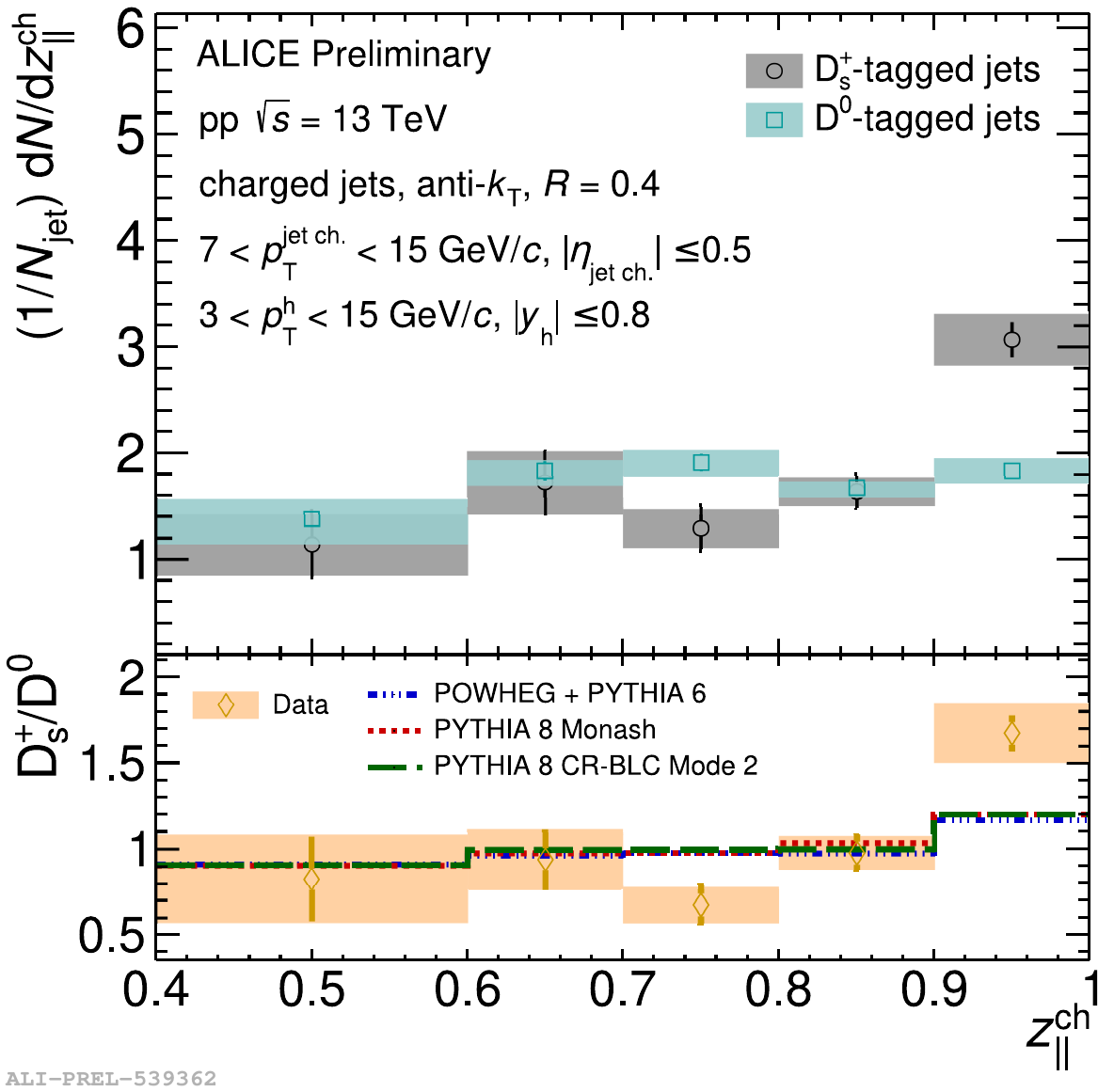} \hspace{2mm}
	\includegraphics[width = 0.46\textwidth]{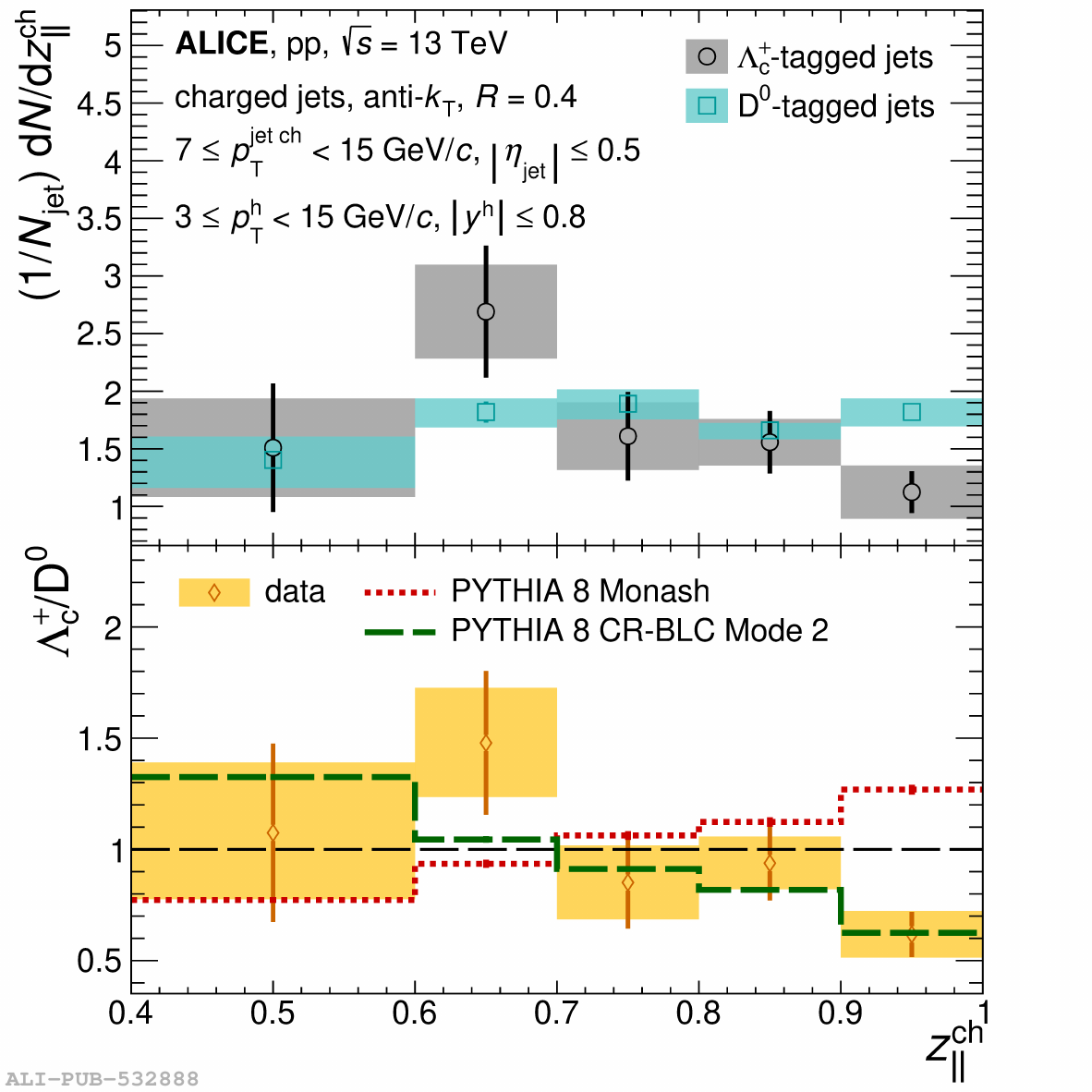}}
	\caption {$\mathrm{D}^{+}_{\mathrm{s}}$ and $\Lambda^{+}_\mathrm{c}$ parallel jet momentum fraction distributions compared to $\mathrm{D^0}$ measurements in the same kinematic region. In the bottom panels, the ratio of the $z_{||}^\mathrm{ch}$ distribution for $\mathrm{D}^{+}_{\mathrm{s}}$ and $\Lambda^{+}_\mathrm{c}$ over $\mathrm{D^0}$, compared to PYTHIA Monte Carlo predictions, is shown.}
	\label{fig:FFs}%
\end{figure}

The hardness of the fragmentation of charm quarks into hadrons can be accessed by investigating the relation between the hadron momentum in the jet direction with respect to the momentum of the charged jets by introducing the parallel jet momentum fraction, defined as:
\begin{equation}
    z_{||}^\mathrm{ch}=\dfrac{\Vec{p}^\mathrm{HF}\cdot\Vec{p}^\mathrm{ch~jet}}{\Vec{p}^\mathrm{ch~jet}\cdot\Vec{p}^\mathrm{ch~jet}}.
\end{equation}
In order to detail the charm-quark fragmentation and to understand possible modifications induced by different hadronisation mechanisms on the charm shower products, the $z_{||}^\mathrm{ch}$ distributions of prompt $\mathrm{D}^{+}_\mathrm{s}$ mesons and $\Lambda^{+}_\mathrm{c}$ baryons in pp collisions at $\sqrt{s}$ = 13 TeV in the kinematic region 7 $< p_\mathrm{T}^\mathrm{ch~j} <$ 15 GeV/$c$ and 3 $< p_\mathrm{T}^{\mathrm{trigger}} < $ 15 GeV/$c$ with $R$ = 0.4 have been measured and compared with previously released $\mathrm{D^0}$ meson-tagged jets measurements~\cite{d0jets}.
An overall good agreement is observed between the $\mathrm{D^{+}_{s}}$ and the $\mathrm{D^0}$ 
parallel jet momentum fraction distributions (Fig.~\ref{fig:FFs}, left), with some tensions for large value of $z_{||}^\mathrm{ch}$, that could be interpreted as the $\mathrm{D}_\mathrm{s}$ carrying, on average, slightly more jet-momentum than the $\mathrm{D^0}$ meson.
The peak at intermediate values of $z_{||}^\mathrm{ch}$ measured for $\Lambda^{+}_\mathrm{c}$-jets (Fig.~\ref{fig:FFs}, right), suggests a possible softer fragmentation of charm quarks in baryons than in mesons.
The comparison with Monte Carlo predictions are represented in the bottom panels of Fig.~\ref{fig:FFs} as the ratio of the $z_{||}^\mathrm{ch}$ distributions, $\mathrm{D}^+_\mathrm{s}$/$\mathrm{D^0}$  and $\Lambda^{+}_\mathrm{c}$/$\mathrm{D^0}$. In the first case, no significant differences are observed between model predictions, overall providing a good description of the data. In the $\Lambda^{+}_\mathrm{c}$/$\mathrm{D^0}$ ratio, the introduction of color reconnection mechanisms beyond leading color, expected to enhance the baryon production, significantly modifies the expected fragmentation function trend with respect to the Monash tune, providing a qualitatively better description of the measurement.

\section{Azimuthal correlations between charm hadrons and charged particles}
Angular correlations of prompt charm hadrons with charged particles in small systems can provide additional details
on the charm fragmentation in small systems with respect to charm-tagged jets.
The typical structure of two-dimensional correlation functions between ``trigger'' charm particles and ``associated'' charged particles is expressed in terms of pseudorapidity and azimuthal angle difference, featuring a near-side (NS) peak at ($\Delta\varphi$, $\Delta\eta$) = (0,0), and an away-side (AS) peak at $\Delta\varphi$ = $\pi$ extending over a wide pseudorapidity range.
The shape and associated particle yields of the near-side peak unveils the internal composition of the charm jet in term of $p_\mathrm{T}$ distribution and multiplicity of its constituents. Moreover, studies of the peak features as a function of the associated charged particle $p_\mathrm{T}$ provide insight into the redistribution of the charm-quark momentum among the other particles produced by its fragmentation in association to the heavy-flavour trigger, as well as their radial displacement from the jet axis.

\begin{figure}[t]
	\centering
	{\includegraphics[width = 0.53\textwidth]{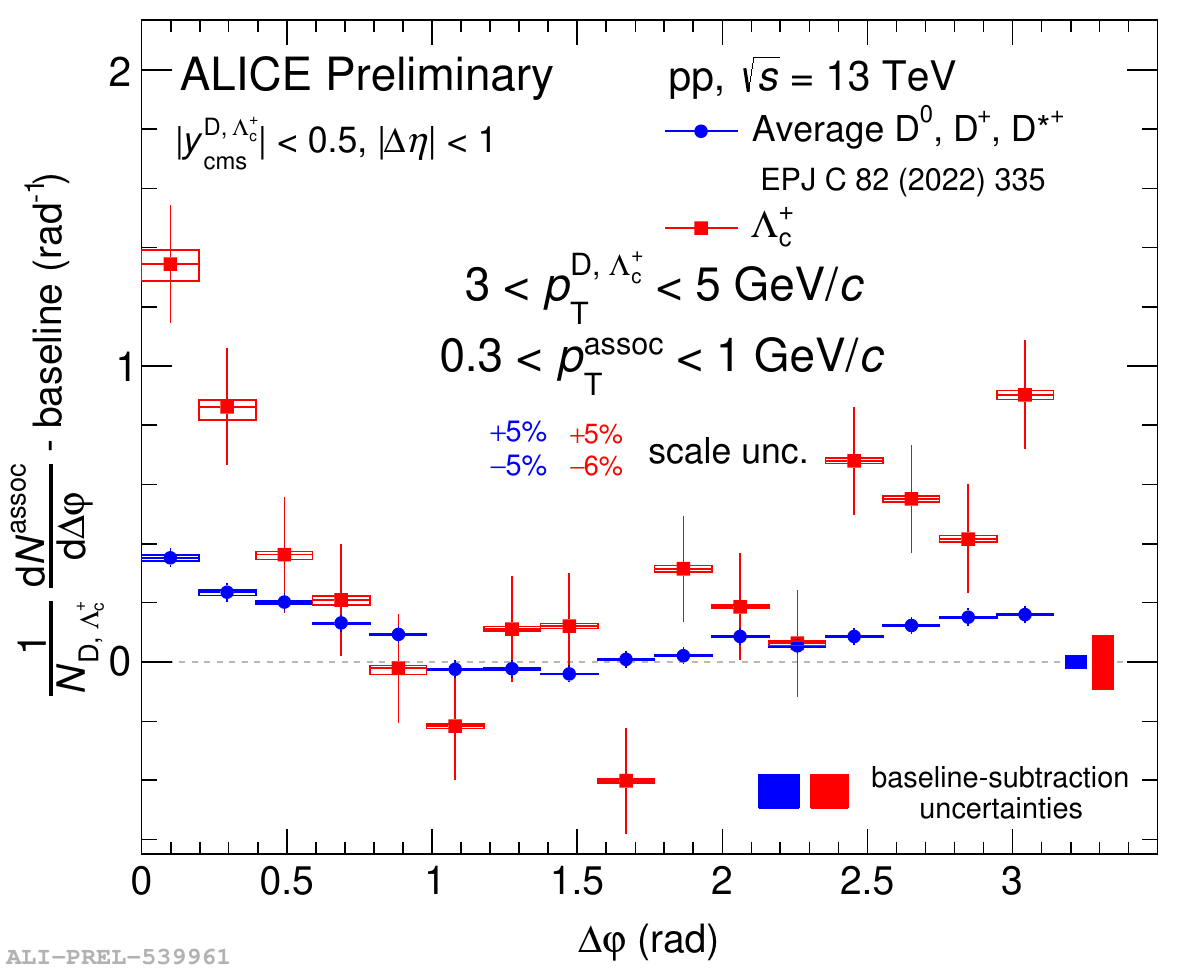} \hspace{2mm}
	\includegraphics[width = 0.39\textwidth]{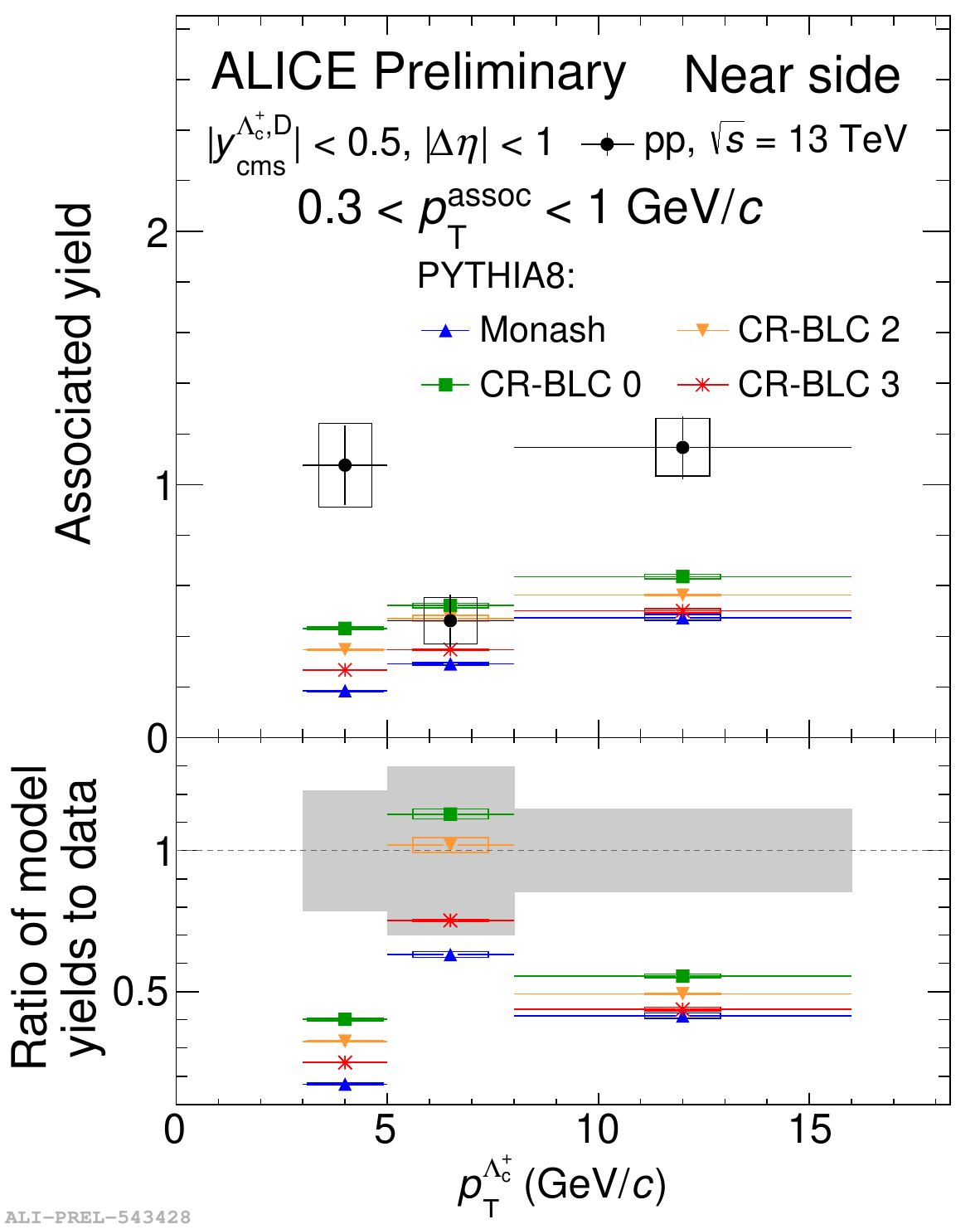}}
	\caption {Left: comparison between $\Lambda^{+}_\mathrm{c}$-hadron and D mesons-hadron azimuthal correlation distributions for 3 $< p_\mathrm{T}^\mathrm{D,\Lambda_\mathrm{c}^{+}} <$ 5 GeV/$c$ and 0.3 $< p_\mathrm{T}^\mathrm{assoc} <$ 1.0 GeV/$c$. Right: comparison between the measured $\Lambda^{+}_\mathrm{c}$-hadron near-side yields and PYTHIA 8 predictions for 0.3 $< p_\mathrm{T}^\mathrm{assoc} <$ 1.0 GeV/$c$.}
	\label{fig:AziLc}%
\end{figure}

$\Lambda^{+}_\mathrm{c}$-hadron azimuthal correlation distributions were measured for the first time in pp collisions at $\sqrt{s}$ = 13 TeV. The comparison with the D meson-charged particles measurements in the same collision system~\cite{Dh@13} can shed light on possible modifications to the charm fragmentation and provide hints about the different hadronisation mechanisms at play when the final state is a meson or a baryon.
A large discrepancy is observed at 3 $< p_\mathrm{T}^\mathrm{D,\Lambda_\mathrm{c}^{+}} <$ 5 GeV/$c$ and 0.3 $< p_\mathrm{T}^\mathrm{assoc} <$ 1.0 GeV/$c$, both in the NS and in the AS region, as shown from the comparison of the azimuthal correlations distribution in Fig.~\ref{fig:AziLc}, suggesting a larger multiplicity of small-momentum associated particle produced, respectively, collimated or in the opposite direction to the $\Lambda^{+}_\mathrm{c}$ baryon.
In addition to different charm hadronisation mechanisms at play, the peaks enhancement could also be possibly explained by assuming a softer fragmentation of the charm baryons than into D mesons, as observed in charm-jet measurements, translating to a larger initial energy of the charm parton, and consequently in more phase space available for the production of other fragmenting particles. Additionally, also the decay of known and unknown heavier charm-baryon states into $\Lambda^{+}_\mathrm{c}$ baryons could possibly cause the enhancement of the NS.
PYTHIA 8 simulations, both with the Monash tune and implementing CR-BLC, are not able to reproduce the $\Lambda^{+}_\mathrm{c}$-hadron distributions, significantly underestimating the NS yields, as illustrated in Fig.~\ref{fig:AziLc} for 0.3 $< p_\mathrm{T}^\mathrm{assoc} <$ 1.0 GeV/$c$. Similarly, also AS yields are underestimated by the Monte Carlo simulations.

\begin{figure}[t]
	\centering
	{\includegraphics[width = 0.640\textwidth]{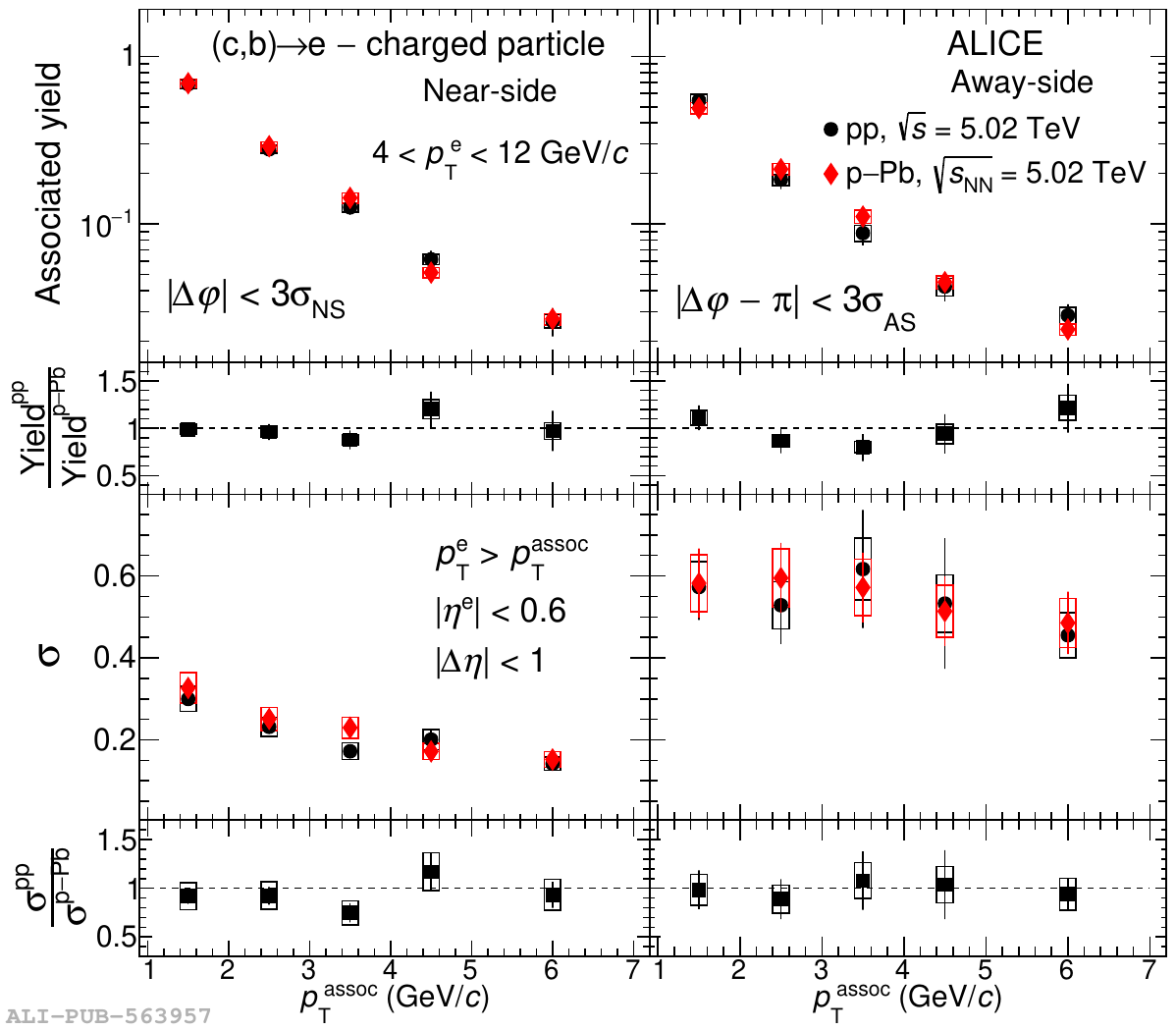}}	
    \caption {Comparison between NS and AS yields (top row) and widths (bottom row) of HF electron-hadron azimuthal correlations in pp and p--Pb collisions.}
	\label{fig:AziEh}%
\end{figure}

More insights about the heavy-flavour quark fragmentation were introduced by studying the azimuthal correlations between electrons from semileptonic decays of HF hadrons and other charged particle in pp and p--Pb collisions at $\sqrt{s_\mathrm{NN}}$ = 5.02 TeV in the integrated transverse momentum range 4 $< p^\mathrm{e}_\mathrm{T} <$ 12 GeV/$c$, shown in Fig.\ref{fig:AziEh}. A good compatibility was observed between the two colliding systems, suggesting negligible impact with current uncertainties of cold nuclear matter effects on the measurements, as it was also observed in D meson-hadron correlation in p--Pb collisions~\cite{Dh5}.

\section{Conclusion}
Charm-jets and azimuthal correlations of charm-origin particles provide a benchmark for studying the fragmentation of charm quarks. Charm-baryon measurements in pp collisions at $\sqrt{s}=$ 13 TeV suggest a softer charm fragmentation with respect to D meson measurements.
While PYTHIA 8 Monash tune, relying on FF constrained to $\mathrm{e^{+}e^{-}}$ and $\mathrm{e^{-}p}$ data, correctly reproduced D mesons measurements, some tensions are observed in the description of $\Lambda^{+}_\mathrm{c}$-jets: including mechanisms sensitive to the surrounding partonic density~(color reconnections) improves the description of the parallel jet momentum fraction distributions, but is not sufficient to explain the $\Lambda_\mathrm{c}$-hadron azimuthal correlation distributions with respect to D-hadron correlations.

The larger data samples collected during the LHC Run 3 data taking and the improved detector performances in tracking and vertexing will allow ALICE to provide more accurate and precise measurements in the charm baryon sector.
Thus, measurements of the observables proposed here as function of the event multiplicity from small to large collision systems will be accessible, enabling the study of possible modifications to the charm fragmentation and hadronisation mechanisms induced by the presence of a deconfined medium.

\end{document}